\documentclass[aps,prl,reprint,showpacs,showkeys,superscriptaddress,preprintnumbers,groupedaddress]{revtex4-1}
\usepackage{amsmath,amssymb,bm,mathrsfs}
\usepackage{srcltx}
\usepackage{dsfont}
\usepackage{epsfig}
\usepackage{slashed}
\usepackage{bbold}
\usepackage{psfrag}
\usepackage{xcolor}
\PassOptionsToPackage{caption=false}{subfig}
\usepackage{subfig}
\usepackage{xfrac}
\usepackage{multirow}
\usepackage{booktabs}
\usepackage[colorlinks=true,linkcolor=blue,citecolor=blue,urlcolor=violet]{hyperref}

\def\({\left(}
\def\){\right)}

\usepackage{soul}

\begin{document}

\preprint{CERN-TH-2017-230}

\title{Catching a New Force by the Tail} 

\author{Simone Alioli}
\email{ simone.alioli@unimib.it}
\affiliation{CERN Theory Division, CH-1211, Geneva 23, Switzerland \& Universita' degli Studi di Milano Bicocca, Piazza della Scienza 3, 20126 Milan, Italy} 

\author{Marco Farina}
\email{farina.phys@gmail.com}
\affiliation{New High Energy Theory Center, Department of Physics, Rutgers University, 136 Frelinghuisen Road, Piscataway, NJ 08854, USA} 

\author{Duccio Pappadopulo}
\email{duccio.pappadopulo@gmail.com}
\affiliation{
Center for Cosmology and Particle Physics,
Department of Physics, New York University, New York, NY 10003, USA.
} 

\author{Joshua T. Ruderman}
\email{ruderman@nyu.edu}
\affiliation{
Center for Cosmology and Particle Physics,
Department of Physics, New York University, New York, NY 10003, USA.
}

\begin{abstract}
The Large Hadron Collider (LHC) is sensitive to new heavy gauge bosons that produce narrow peaks in the dilepton invariant mass spectrum up to about $m_{Z'}\sim 5$~TeV\@.  $Z'$s that are too heavy to produce directly can reveal their presence through interference with Standard Model dilepton production.  We show that the LHC can significantly extend the mass reach for such $Z'$s by performing precision measurements of the shape of the dilepton invariant mass spectrum. The high luminosity LHC can exclude, with 95\% confidence, new gauge bosons as heavy as $m_{Z'} \sim 10-20$~TeV that couple with gauge coupling strength of $g_{Z'} \sim 1-2$. 
 \end{abstract}

\maketitle

{ \em Introduction.---} Apart from gravity and the Higgs force, all known forces are mediated by spin-1 particles: the photon for electromagnetism, the $W/Z$ bosons for the weak force, and gluons for the strong force.

The search for new forces and their massive mediators is a well-motivated arena for both experiment and theory. New short range abelian gauge forces appear in many extensions of the Standard Model (SM)~\cite{Langacker:1984dc,London:1986dk,Altarelli:1989ff,Leike:1998wr,Erler:2002pr,Dittmar:2003ir,Carena:2004xs,Freitas:2004hq,Rizzo:2006nw,Chankowski:2006jk,Wells:2008xg,Rizzo:2009pu,Li:2009xh,Erler:2009jh,Diener:2009vq,delAguila:2010mx,Erler:2011ud,Chiang:2011kq,Han:2013mra,Accomando:2015cfa,Fuks:2017vtl,Greljo:2017vvb} (see also~\cite{Langacker:2008yv,Patrignani:2016xqp} for reviews), are an active area of investigation at the LHC~\cite{Aad:2012hf,Chatrchyan:2012oaa,Aad:2014cka,Khachatryan:2014fba,Khachatryan:2016zqb,Aaboud:2016cth,Aaboud:2017buh}, and serve as standard benchmarks to test the performances of future colliders~\cite{CMS:2013xfa,ATLAS:2013hta,Hayden:2013sra,Godfrey:2013eta,Gershtein:2013iqa,Pappadopulo:2014qza,Thamm:2015zwa,Golling:2016gvc}.
Additional non-anomalous $U(1)$ gauge groups ~\cite{Appelquist:2002mw,Ferroglia:2006mj,Salvioni:2009mt,Salvioni:2009jp,Basso:2011na, Basso:2008iv,Basso:2010jt,Basso:2010pe,Basso:2010jm,Basso:2010yz} are a relatively innocuous extension of the SM as the masses of the associated vector bosons do not require the existence of additional scalar degrees of freedom and consequently, a worsening of the hierarchy problem.

The traditional strategy to search for $Z'$s at colliders has been to perform ``bump hunts."  For $Z'$s decaying to leptons, the dilepton invariant mass distribution is scrutinized for narrow peaks rising above the monotonically falling background.  
Searches at the LHC are sensitive to $Z's$ with masses up to about 5 TeV~\cite{Aad:2012hf,Chatrchyan:2012oaa,Aad:2014cka,Khachatryan:2014fba,Khachatryan:2016zqb}.  

\begin{figure}[!!!t]
\begin{center}
\includegraphics[width=0.4 \textwidth]{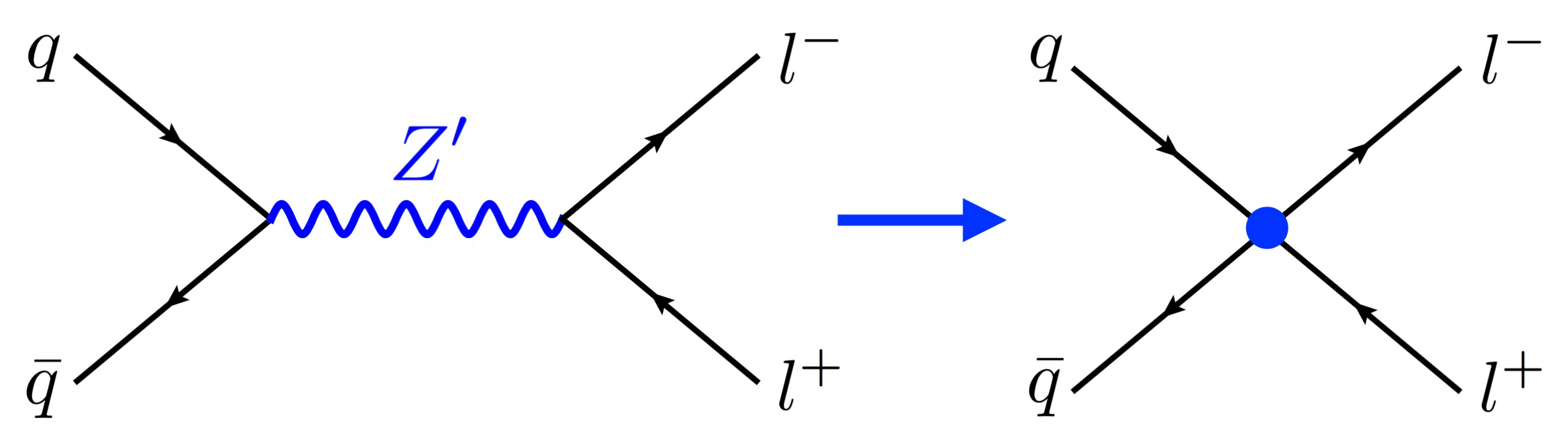}
\end{center}
\vspace{-.3cm}
\caption{\small \it At energies $E$ much smaller than the mass $M$ of the heavy gauge boson $Z'$, the effect of the new physics on the Drell-Yan process, $pp\to\ell^+\ell^-$, is encoded by a finite set of four-fermion contact operators.
 \label{schema}}
\end{figure}

For masses above 5 TeV, bump hunts lose sensitivity as the cross section for direct production vanishes. When the mass $M$ of the new vector boson is too large for direct production, the main contribution of the $Z'$ at energies $E\ll M$ are interference effects~\cite{Dittmar:1996my,Petriello:2008zr,Accomando:2010fz,Accomando:2013sfa}, which modify the shapes of kinematical distributions. If the $Z'$ couples to both quarks and  leptons, it modifies the invariant mass distribution of Drell-Yan processes $pp\to \ell^+\ell^-$, $\ell=e,\mu$. 
The interference effects can be captured by a small number of higher dimension operators, obtained by integrating out the $Z'$ (see Fig.~\ref{schema}), and are therefore relatively insensitive to the specific details of the $Z'$ model.

In this letter, we assess the reach of the LHC to probe heavy $Z's$ through precision fits to the shape of the invariant mass spectrum of dileptons.   Previous studies of the interference of heavy Z's at the LHC found that a 5 sigma discovery will be difficult~\cite{Rizzo:2009pu}, and estimated the reach of early 13 TeV measurements~\cite{Greljo:2017vvb}.  We go beyond these preliminary studies by performing the first  comprehensive study of theoretical uncertainties and their correlations, and by mapping the future reach of the full LHC dataset.  We find that a vast parameter space of Z's will be probed at the LHC\@.  Deviations in the shape of the Drell-Yan distribution have also been used to constrain effective operators~\cite{Farina:2016rws}, the running of electroweak gauge couplings~\cite{Rainwater:2007qa,Alves:2014cda}, and other radiative effects of new electroweak states~\cite{Matsumoto:2017vfu}.

The rest of this letter is organized as follows.  We begin by reviewing the class of $Z'$ models that we study.  Then we present the reach we find of the LHC to the interference effects of heavy $Z's$.  We finish with our conclusions.  We include appendices that contain a technical description of our SM prediction, projections with future higher energy colliders, and a comparison of our bounds with  experimental contact operator bounds.

\begin{figure*}[!!!t]
\begin{center}
\includegraphics[width=1 \textwidth]{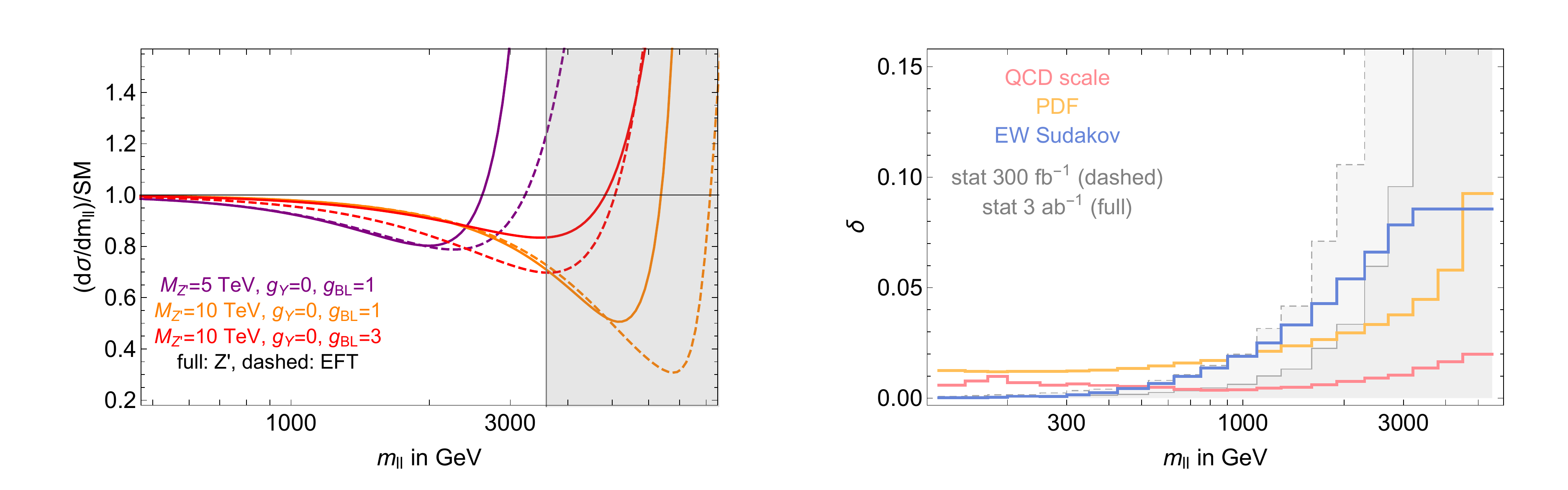}
\end{center}
\vspace{-.3cm}
\caption{\small \it
{\bf{Left panel}}: Ratio of the dilepton invariant mass distribution in the $Z'$ model to the SM\@. The solid lines are calculated using the full model of Eq.~\ref{Zprimeint}, while the dashed lines are calculated using the EFT of Eq.~\ref{Zprimeeft}. In the gray region, there are 3 expected SM events with a luminosity of 3\,ab$^{-1}$. 
{\bf{Right panel}}: Systematic theoretical uncertainties used in our analysis. We also show the size of the statistical uncertainty associated to the SM prediction.
 \label{distributions}}
\end{figure*}

{ \em The Minimal Model.---} A class of $Z'$ models motivated by their simplicity and minimality has been studied in~\cite{Appelquist:2002mw,Ferroglia:2006mj,Salvioni:2009mt,Salvioni:2009jp,Basso:2011na,Basso:2008iv,Basso:2010jt,Basso:2010pe,Basso:2010jm,Basso:2010yz}. These Minimal $Z'$ Models are defined by the requirement that the new $U(1)$ vector boson gauges a linear combination of the hypercharge ($Y$) and the difference between baryon and lepton number ($B-L$) currents. This ensures that the model is anomaly free as long as right-handed neutrinos are present. The gauge structure also ensures flavor universal interactions for the new vector field.

The Lagrangian describing the interactions of the new vector boson $\mathcal Z$ can be written as
\begin{equation}\label{Zprimeint}
\mathscr L\!=\!-\frac{1}{4}\mathcal Z_{\mu\nu}^2+\frac{M^2}{2}\mathcal Z_\mu^2-\mathcal Z_\mu (g_YJ_H^\mu+g_Y J^\mu_Y+g_{BL}J^\mu_{BL}) \, ,
\end{equation}
where  $J_Y^\mu=\sum_f Q^{(f)}_Y\bar f\gamma^\mu f$ and $J_{BL}^\mu=\sum_f Q^{(f)}_{BL}\bar f\gamma^\mu f$ are the fermionic hypercharge and $B-L$ currents, respectively, and $J_H^\mu\equiv i Q^{(H)}_Y (H^\dagger D^\mu H -  D^\mu H^\dagger H)$. The SM field charges $Q_Y$ and $Q_{BL}$ are shown in Table~\ref{charges}. The couplings $g_Y$ and $g_{BL}$ define the strength of the interactions between the $\mathcal Z$ boson and the respective currents.

The spectrum contains three neutral vector bosons: a massless photon and two massive vectors, to be identified with the $Z$ boson and the heavy $Z'$. 
When $g_Y\neq 0$, the coupling between $\mathcal Z$ and the Higgs boson current leads to a mixing between the $Z$ boson and $\mathcal Z$.
 Their masses are approximately given by $m_Z\approx g_Z v/2\equiv  m_{Z_0}$ and $\quad m_{Z'}\approx M$ with $g_Z^2\equiv g'^2+g_2^2$ and $v=246$\,GeV\@. Corrections to this equations are small, of order $(g_Y^2/g_Z^2)(m_{Z_0}^2/M^2)$, which is also the typical size of the corrections to electroweak observables.
In terms of the gauge eigenstates $B$, $W_3$, and $\mathcal Z$,
\begin{equation}
Z=\cos\alpha  Z_0-\sin\alpha \mathcal Z,\quad Z'=\sin\alpha Z_0+\cos\alpha  \mathcal Z \, ,
\end{equation}
where $Z_0$ is the unperturbed $Z$ boson wave function $Z_0\propto g_2 W_3- g' B$ and
\begin{equation}
\tan 2\alpha =\frac{2g_Y/g_Z   \,m_{Z_0}^2}{M^2-m_{Z_0}^2(1- g_Y^2/g_Z^2)}\approx 2\frac{g_Y}{g_Z}\frac{m_{Z_0}^2}{M^2}.
\end{equation}

\begin{table}[t]
\renewcommand{\arraystretch}{1.}
{\small
\begin{tabular}{c|cccccc} 
$f$ & $H$&$\ell_L $& $e_R$& $q_L$& $u_R$& $d_R$\\ \hline \hline
$Q_Y$ & $1/2$&$-1/2 $& $-1$& $1/6$& $2/3$& $-1/3$\\
$Q_{BL}$ & $0$&$-1 $& $-1$& $1/3$& $1/3$& $1/3$
\end{tabular}
} 
\caption{\label{charges}\it Hypercharge and $B-L$ charges.}
\end{table}

The coupling of the physical vector bosons to SM fermions are
\begin{equation}
J_Z^\mu =\cos\alpha J^\mu_{Z_0}-\sin\alpha J_{\mathcal Z}^\mu,\quad J_{Z'}^\mu =\sin\alpha J^\mu_{Z_0}+\cos\alpha J_{\mathcal Z}^\mu
\end{equation}
where $J_{Z_0}^\mu$ is the $Z$ boson current in the SM, $J^\mu_{Z_0}=g_Z\sum_f \bar f\gamma^\mu(T_{3L}-\sin^2\theta_W Q) f$, and $J_{\mathcal Z}^\mu=g_Y J^\mu_Y+g_{BL}J^\mu_{BL}$.

At energies $E\ll M$ the physics described by Eq.~\ref{Zprimeint} is captured by an Effective Field Theory (EFT) obtained by integrating out $\mathcal Z$. At leading order in $1/M$ this is given by
\begin{equation}\label{Zprimeeft}
 \mathscr L_{EFT} =-\frac{1}{2 M^2}\left(g_Y J_H^\mu+g_Y J_Y^\mu+g_{BL} J_{BL}^\mu\right)^2.
\end{equation}

In the left panel of Fig.~\ref{distributions} we show the ratio of the dilepton invariant mass distribution in the presence of a $Z'$ to the SM, and compare the results obtained from the full theory in Eq.~\ref{Zprimeint} to the EFT of Eq.~\ref{Zprimeeft}.  
The two calculations agree for invariant masses within the reach of the LHC, when the $Z'$ is heavy and not too wide. 

\begin{figure}[!!!t]
\begin{center}
\includegraphics[width=0.42 \textwidth]{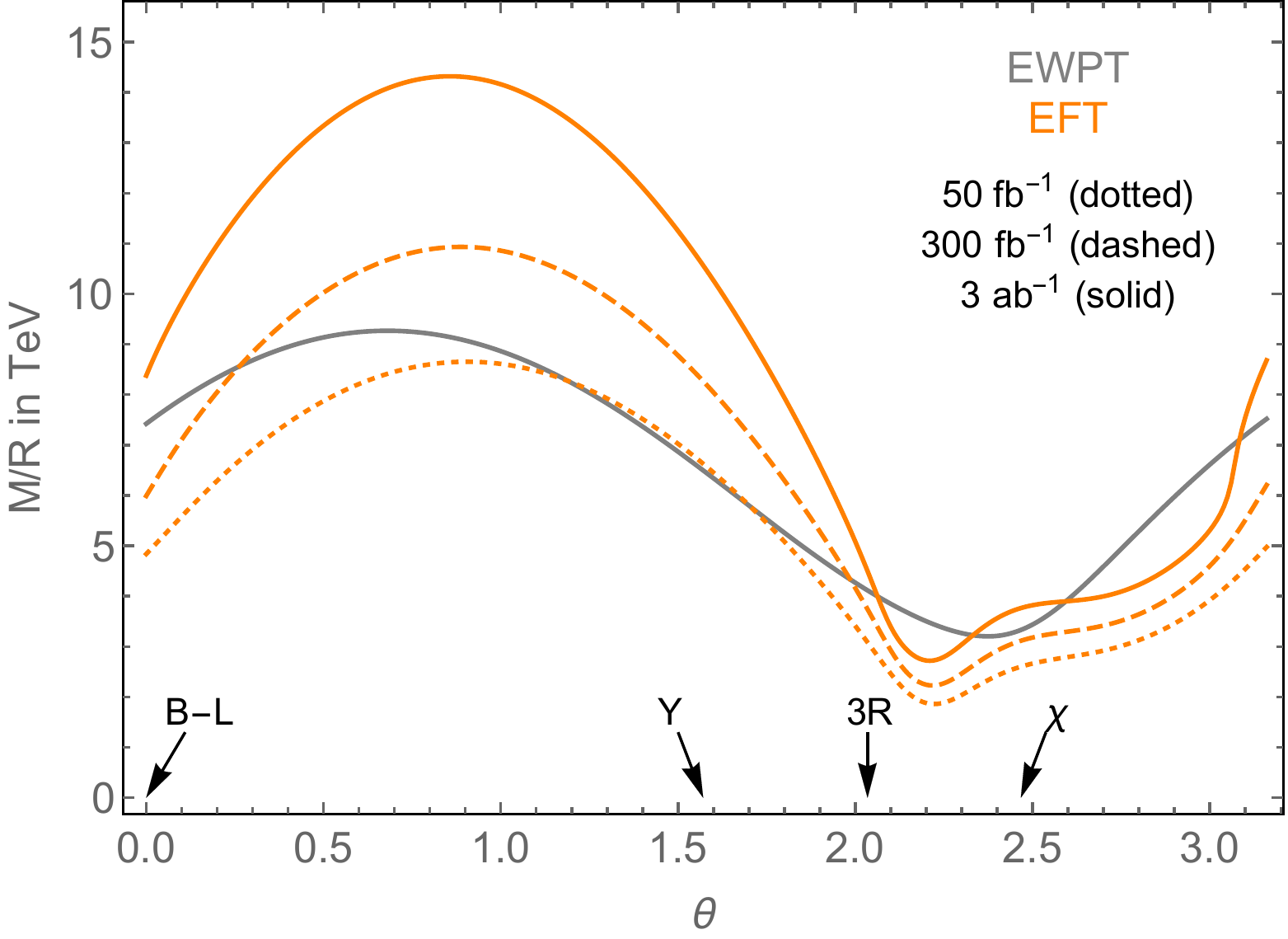}
\end{center}
\vspace{-.3cm}
\caption{\small \it
95\% C.L. lower bound on $M/R$ as a function of $\theta$. The constraint coming from low energy experiments is shown in gray, while the reach of the LHC is shown in orange. Specific models are identified: pure B-L ($\tan\theta=0$), hypercharge ($\tan\theta=+\infty$), $T_{3R}$ ($\tan\theta$=$-2$), and $U(1)_\chi$ ($\tan\theta$=$-4/5$)~\cite{Patrignani:2016xqp,Langacker:2008yv}. Notice that these relations are defined at tree level.
 \label{geff}}
\end{figure}

{ \em Existing bounds and projections.---} In our analysis we consider two kinds of constraints on Minimal $Z'$ Models. The first set comes from low energy measurements,  including constraints  from LEPI and LEPII~\cite{Barbieri:2004qk,Cacciapaglia:2006pk}. These can be evaluated using the low energy Lagrangian in Eq.~\ref{Zprimeeft}, and depend on the parameter combinations $g_Y/M$ and $g_{BL}/M$. 
We extract these bounds from the global fit in~\cite{Falkowski:2017pss}. 

The second set of constraints comes from the LHC measurements of the dilepton invariant mass distribution in $pp\to \ell^+\ell^-$ collisions. If the $Z'$ is light enough to be produced on-shell, it will manifest as a resonant excess in the dilepton spectrum. Bump hunt searches are optimized to look for this kind of isolated excess \cite{Aad:2012hf,Chatrchyan:2012oaa,Aad:2014cka,Khachatryan:2014fba,Khachatryan:2016zqb}. The results presented in \cite{Khachatryan:2016zqb} are currently the strongest constraints on Minimal $Z'$s for $M < 5$\,TeV\@. 

In this section we establish the reach of the LHC for $Z'$s that are too heavy to be efficiently produced on-shell, and thus escape bump hunt searches. A $Z'$ will distort the high energy tail of the dilepton invariant mass distribution, as shown in the left panel of Fig.~\ref{distributions}. 
To project the sensitivity to Z's, we must to predict both the SM and the New Physics contributions to the dilepton spectrum.  

For the SM prediction, we use Next-to-Next-to-Leading Order (NNLO) QCD~\cite{Hamberg:1990np,Anastasiou:2003yy,Anastasiou:2003ds,Melnikov:2006kv,Catani:2009sm,Gavin:2010az} and Next-to-Leading Order (NLO) ElectroWeak (EW)~\cite{Berends:1984xv,Baur:1997wa,Baur:2001ze,CarloniCalame:2007cd,Arbuzov:2007db,Dittmaier:2009cr}  results from  {\tt{FEWZ-3.1}}~\cite{Li:2012wna}. Details about the generation and the evaluation of the QCD and Parton Distribution Function (PDF) uncertainties are reported in the Appendix.  The NLO EW corrections in FEWZ include virtual $\gamma,Z$, and $W$ exchange and real QED corrections, but do not include real $W$ or $Z$ emissions.  Real $W/Z$ emissions could be important at high energies, in an inclusive measurement, and we include them after calculating them separately at leading order using {\tt{MadGraph5\_aMC@NLO}}~\cite{Alwall:2014hca} (see~\cite{Baur:2006sn} for a previous calculation).

Our treatment of EW uncertainties, which is described in the Appendix, is designed to capture the effect of missing two-loop Sudakov logarithms~\cite{Baur:2006sn,Bell:2010gi,Becher:2013zua,Ciafaloni:2000df,Ciafaloni:2000rp,Ciafaloni:2000gm,Ciafaloni:2001vt,Ciafaloni:2000df,Ciafaloni:2000rp,Ciafaloni:2000gm,Ciafaloni:2001vt,Manohar:2014vxa}. 
A visual summary of the various theoretical uncertainties is presented on the right panel of Fig.~\ref{distributions}.

New Physics predictions are calculated at Leading Order (LO) and are multiplied by the ratio of the SM NNLO QCD cross section to the SM LO one.
This is justified by the fact that the relative size of the NNLO QCD corrections only depend on the invariant mass of the dilepton system. Both the SM and New Physics cross sections are calculated in the dilepton invariant mass bins shown in the right panel of Fig.~\ref{distributions}. We apply a $p_T > 25$\,GeV and $| \eta| < 2.5$ cut on leptons, and we assume 65\% (80\%) identification efficiency for di-electron (di-muon) events, motivated by past LHC Drell-Yan measurements~\cite{CMS:2014jea,Aad:2016zzw}.

\begin{figure*}[!!!t]
\begin{center}
\includegraphics[width=1 \textwidth]{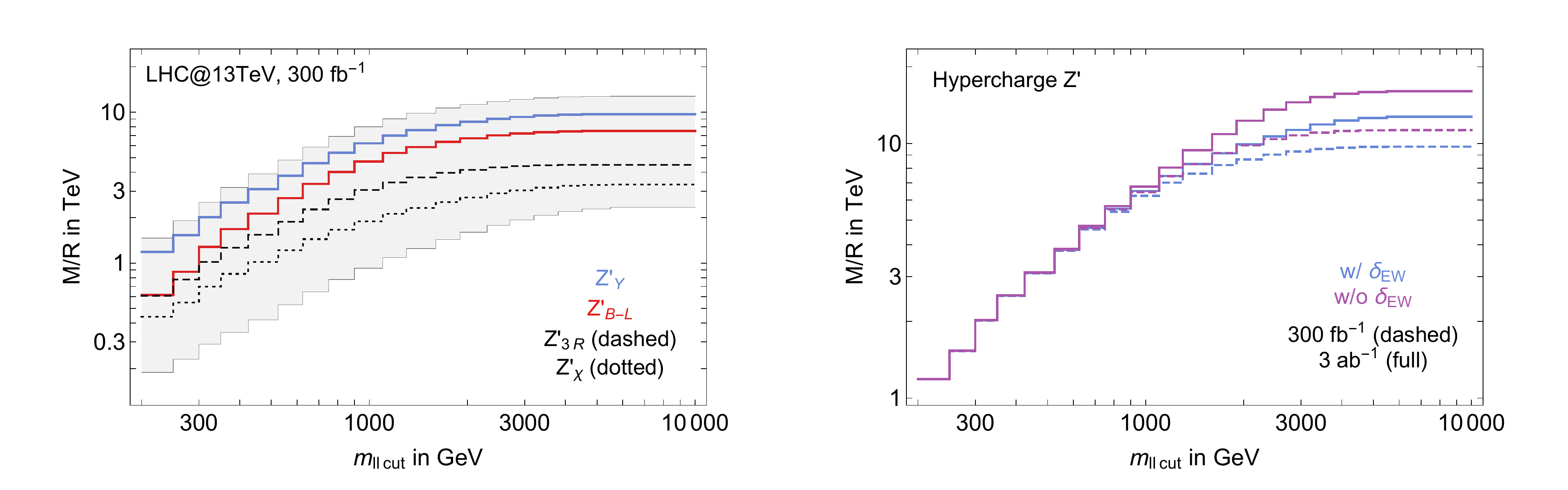}
\end{center}
\vspace{-.3cm}
\caption{\small \it
{\bf{Left panel}}: 95\% CL lower bound on $M/R$ as a function of $m_{\ell\ell\,{\textrm{cut}}}$, for three example models, defined by specific choices of $\theta$ (see Fig.~\ref{geff}). {\bf{Right panel}}: 95\% CL lower bound on $M/R$ for the hypercharge model ($\theta=\pi/2$) as a function of $m_{\ell\ell\,{\textrm{cut}}}$. We show how the bound differs using two different choices for the total integrated luminosity ($300$\,fb$^{-1}$ and 3\,ab$^{-1}$) and switching off the theoretical uncertainty on higher order EW corrections. 
 \label{mcutplot}}
\end{figure*}

In order to infer the reach for Minimal $Z'$ Models, we fit to the Born-level cross section, after unfolding detector effects.  We perform a $\chi^2$ test, including QCD-scale, EW, and PDF uncertainties, and their respective correlations across different bins of invariant mass.  The scale and EW uncertainties are fully correlated, whereas the PDF uncertainties exhibit nontrivial correlations. The Appendix describes our $\chi^2$ test, and the derivation of PDF correlations, in more detail.
There is also an experimental uncertainty due to unfolding, arising from effects such as detector resolution, energy scale, and lepton identification efficiency.  We estimate the experimental uncertainty as being composed of uncorrelated and fully correlated components, both of which we take to be 5\% of the cross section, bin-by-bin.  This choice is motivated by the size of experimental uncertainties in previous unfolded Drell-Yan measurements conducted at $\sqrt s = 8$~TeV~\cite{CMS:2014jea,Aad:2016zzw}.

Fig.~\ref{geff} shows the comparison between the low energy bounds and projected LHC bounds on the Minimal $Z'$ Models extracted using the EFT of Eq.~\ref{Zprimeeft}. We introduce an angular variable, $\tan\theta\equiv g_{Y}/g_{BL}$, and rewrite Eq.~\ref{Zprimeeft} as a function of the dimensionless coefficient $R\equiv\sqrt{g_Y^2+g_{BL}^2}/g_Z$. For a given value of $\theta$, the 95\% C.L. lower bound $M/R$ is shown in Fig.~\ref{geff}. Surprisingly, the LHC starts to be competitive with low energy bounds around the present time, with an integrated luminosity of 50\,fb$^{-1}$. 

Notice that while the bounds extracted from the low energy experiments have a wide range of applicability, in terms of the mass $M$ of the resonance, the LHC bounds require the resonance to be heavy enough so that Eq.~\ref{Zprimeeft} can be used to describe the Drell-Yan process in a particular invariant mass bin. In order to obtain bounds which are applicable for a variety of masses, we adopt the following procedure~\cite{Farina:2016rws,Alioli:2017jdo}. We recalculate the projected 95\% CL upper bound on $M/R$ by including 
only those invariant mass bins for which $m_{\ell\ell}< m_{\ell\ell\,{\textrm{cut}}}$. For a given $Z'$ of mass $M$, a consistent bound on its coupling $R$ is obtained using $m_{\ell\ell\,{\textrm{cut}}}=\alpha M$, where $\alpha\lesssim 1$. $\alpha$ can in principle depend on the width of the resonance, $\Gamma_{Z'}$. The bounds on $M/R$ as a function of $m_{\ell\ell\,{\textrm{cut}}}$ are shown in Fig.~\ref{mcutplot}. For small $m_{\ell\ell\,{\textrm{cut}}}$, the bound is weaker, while it saturates for $m_{\ell\ell\,{\textrm{cut}}}\gtrsim 3$\,TeV, above which the energy growth of the partonic cross section is counterbalanced by the decrease of the parton luminosities. The right panel of Fig.~\ref{mcutplot} shows that an increase in the total integrated luminosity strengthens the bound only for $m_{\ell\ell\,{\textrm{cut}}}\gtrsim 1$\,TeV, since in this case the total uncertainty is dominated by the statistical uncertainty.

\begin{figure}[!!!b]
\begin{center}
\includegraphics[width=0.42 \textwidth]{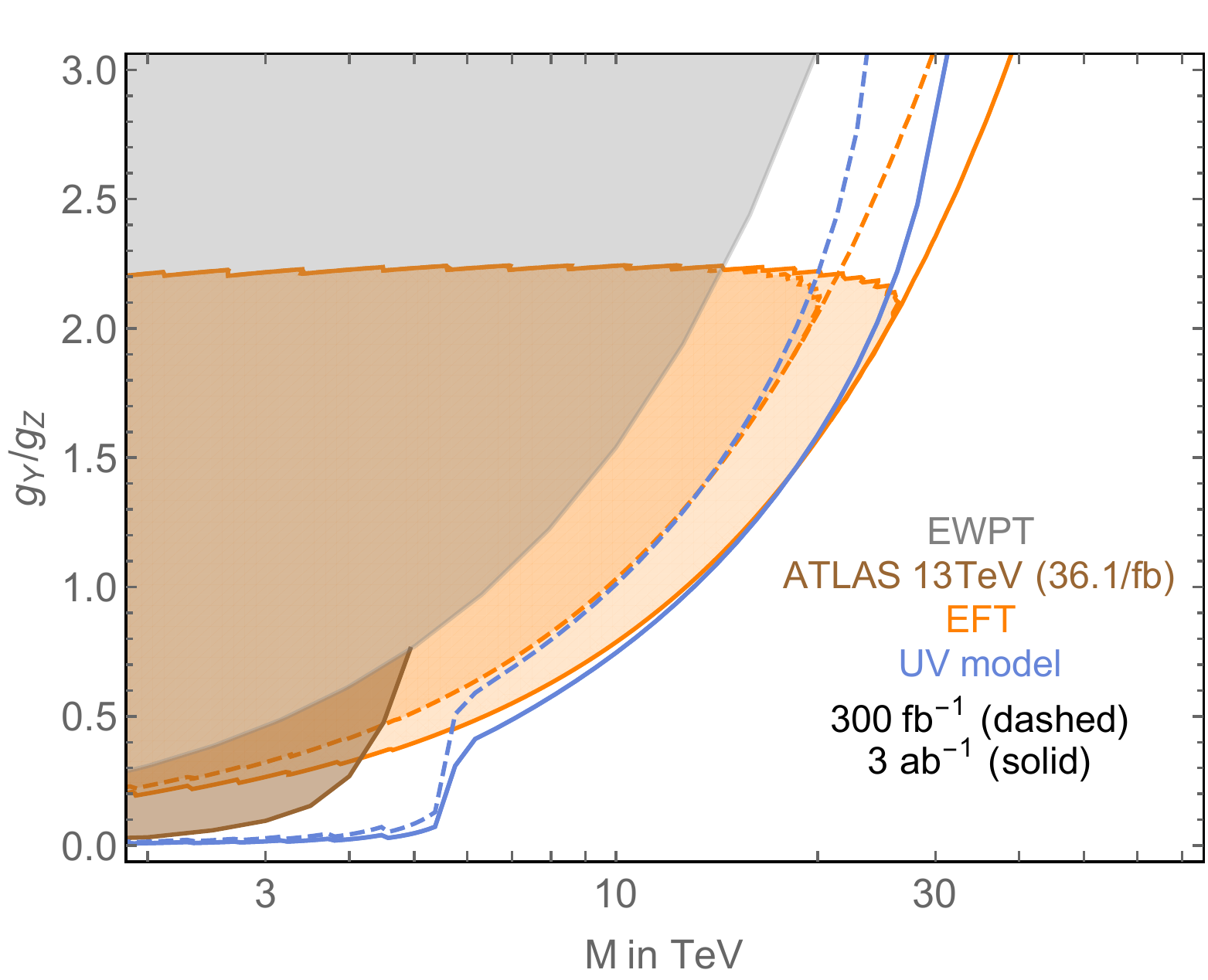}
\end{center}
\vspace{-.3cm}
\caption{\small \it
Comparison between the 95\% CL upper bound on $g_Y$ extracted using the EFT of Eq.~\ref{Zprimeeft}, with $m_{\ell\ell\,{\textrm{cut}}}=M-2.5\times\Gamma_{Z'}$, and the full model Eq.~\ref{Zprimeeft}. The two bounds agree for masses $5.5\,{\textrm{TeV}}\lesssim M\lesssim25\,{\textrm{TeV}}$. For smaller $M$, the EFT does not capture on-shell $Z'$ production and the bound extracted from the full model is much stronger. At larger masses and couplings, finite $Z'$ width effects, which are not included in the EFT calculation, become important and  lead to a weakening of the bound in the full model. The gray region shows the region which is excluded by low energy measurements.
 \label{uvplot}}\end{figure}

It is  natural to ask how the bounds on a given $Z'$ model, obtained from the full theory in Eq.~\ref{Zprimeint}, compare with those extracted from the EFT of Eq.~\ref{Zprimeeft}.
Using the hypercharge model as a benchmark, Fig~\ref{uvplot} shows the  95\% CL upper bound on the coupling $g_Y$,
using the full model in Eq.~\ref{Zprimeint}. 
We compare to the exclusion obtained from the EFT, where we choose either $m_{\ell\ell\,{\textrm{cut}}}=\infty$ or $m_{\ell\ell\,{\textrm{cut}}}=M-2.5\times\Gamma_{Z'}$. 

Fig.~\ref{uvplot} shows that for small enough $M \lesssim 5.5$\,TeV, the EFT bound is much weaker than the one obtained from the full model. In this region, the cross section is dominated by on-shell $pp\to Z'$ production, followed by $Z'\to\ell^+\ell^-$ decay. 
 The bound in this region approximates the reach of bump hunt searches, and we find a result consistent, within a factor of 2 in cross section, to prior bump-hunt studies~\cite{Godfrey:2013eta,Thamm:2015zwa}.
At larger masses, the bound on $g_Y$ agrees when using the full model versus the EFT\@.  The agreement stops around $M\sim25$\,TeV and $g_Y/g_Z\gtrsim 2.5$.
 At large coupling, the $Z'$ width is correspondingly larger and $\Gamma_{Z'}/M$ corrections become important. These lead to a cancellation in the size of the deviation from the SM prediction (see the red curve in Fig~\ref{distributions}).

 Here we have focused on $2\sigma$ exclusions.  When  $M \gtrsim 5.5~\mathrm{TeV}$, we find that a $5\sigma$ discovery is not possible at at the LHC, given LEP bounds.   However it is possible to have a signal with $3\sigma$ significance.  Additional 95\% C.L. projections for a $pp$ collider with a larger center of mass energy (27 and 100\,TeV) are shown in the Appendix.


{ \em Conclusions.---}In this letter we have shown that precision measurements of the shape of the dilepton invariant mass spectrum have broad reach to probe off-shell $Z'$s, extending the mass reach of direct searches.  Unlike bump hunts, off-shell interference is insensitive to the presence of other decay modes.  Our results only rely on the invariant mass distribution, but it would be interesting to explore how much sensitivity is gained by also using angular information.  We have demonstrated significant reach for $Z'$s, after a careful accounting of theoretical uncertainties.  In order to fully realize this reach, our results motivate a concerted effort to control experimental uncertainties in energetic dilepton tails.  The LHC may retain significant power, even if new physics is too heavy for direct production.

\vspace{.3cm}
\begin{acknowledgements}
{\em Acknowledgements.---} We thank Bogdan Dobrescu, Stefano Forte, Frank Petriello, Eram Rizvi, Tom Rizzo, and Hwidong Yoo for helpful discussions.  We thank Christian Bauer for providing code that calculates the resummed EW corrections from Ref.~\cite{Bauer:2016kkv}.  SA acknowledges support
by the COFUND Fellowship under grant agreement PCOFUND-GA-2012-600377.
MF is supported in part by the DOE Grant DE-SC0010008. DP and JTR are supported by NSF CAREER grant PHY-1554858. This work was supported in part
by the hospitality of the Aspen Center for Physics, which is supported by National Science
Foundation grant PHY-1066293.
\end{acknowledgements}
\appendix

\section{Appendices}
{ \em SM predictions.---}The SM prediction for the binned dilepton invariant mass distribution at 13, 27, and 100 TeV are calculated using {\tt{FEWZ-3.1}}~\cite{Li:2012wna}. This provides NNLO QCD accuracy and part of the NLO electroweak corrections, including the photon-initiated channel. Scale and PDF uncertainties are calculated using a 7-point envelope $(\mu_r,\mu_f) = (1,1),(2,1),(2,2),(0.5,1),(0.5,0.5),(1,2),(1,0.5) \times M_{\ell\ell}$ and the 107 symmetric eigenvectors of the {\tt{LUXqed\_plus\_PDF4LHC15\_nnlo\_100}} PDF set \cite{Butterworth:2015oua,Manohar:2016nzj} with $\alpha_s(m_Z)=0.118$.

\begin{figure}[!!!t]
\begin{center}
\includegraphics[width=0.42 \textwidth]{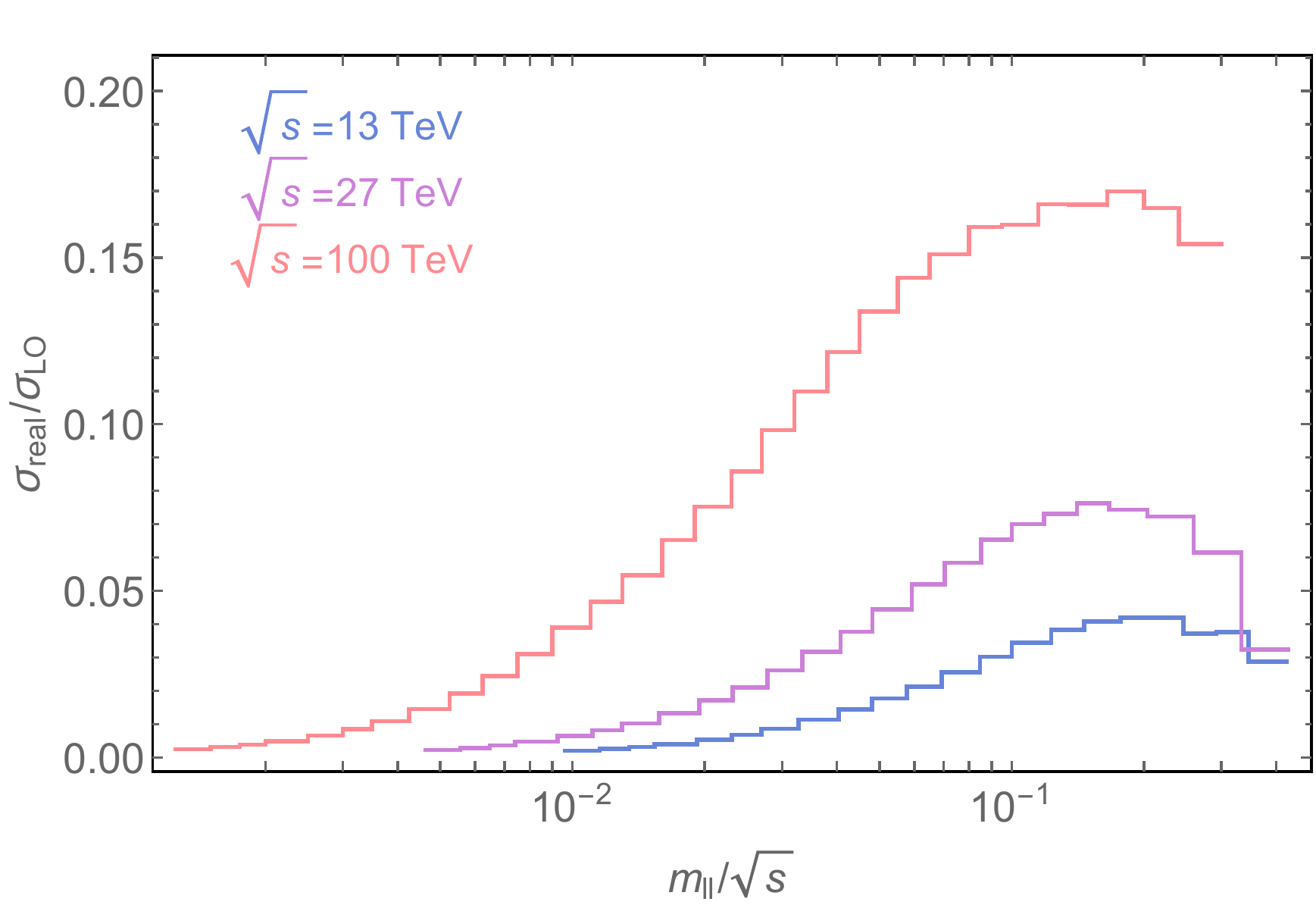}
\end{center}
\vspace{-.3cm}
\caption{\small \it
Relative correction to the LO Drell-Yan cross section due to real $W^{\pm}/Z$ emission, calculated at 13, 27, and 100\,TeV center of mass energy.
 \label{realemission}}\end{figure}

We include corrections from real $W$ and $Z$ radiation, which we evaluate at LO using {\tt{MadGraph5\_aMC@NLO}} \cite{Alwall:2014hca}. Their fractional size, with respect to the LO Drell-Yan cross section, is shown in Fig.~\ref{realemission}.

The size of higher order EW effects, which are not included by {\tt{FEWZ}}, must be estimated and included as a systematic uncertainty. The leading missing effect corresponds to two-loop EW Sudakov logarithms, representing a fractional correction to the Drell-Yan cross section of order $\alpha^2_{EW}/(4\pi^2)\log^4\sqrt s/m_W^2$. {\tt{FEWZ}} provide two choices of EW input scheme and one could try to estimate the size of the EW uncertainty comparing the Drell-Yan cross section in these two schemes. We notice that doing this would not capture the leading effect. In order to estimate the contribution of two-loop EW Sudakov logarithms, we use the results of \cite{Bauer:2016kkv}. As a function of the partonic center of mass energy of the $pp\to \ell^+\ell^-$ process, \cite{Bauer:2016kkv} shows the fractional difference between the fixed order and the resummed calculation (see for instance Fig.~3 in \cite{Bauer:2016kkv}). We use this difference to estimate the size of the two-loop Sudakov logarithm.  Note that the results of \cite{Bauer:2016kkv} depend on the partonic center of mass energy, while we are interested in the dilepton invariant mass. These two variables are equivalent in the limit of soft radiation, which we assume in order to use the result of \cite{Bauer:2016kkv}.
The size of the EW uncertainty calculated in this way is shown in Fig.~\ref{distributions}.

In order to infer the reach for the Minimal $Z'$ Models, we use a $\chi^2$ test,
\begin{equation}\label{llh}
\chi^2= \sum_{i,j}(\sigma_i-\sigma_i^{SM})(\Sigma^{-1})_{ij}(\sigma_j-\sigma_j^{SM}),
\end{equation}
where $\sigma_i$ and $\sigma_i^{SM}$ are the cross sections in the $i$-th invariant mass bin for the $Z'$ model and the SM respectively. $\Sigma$ is the full covariance matrix, calculated as the sum $\Sigma=\Sigma^{\rm{stat}}+\Sigma^{\rm{th}}+\Sigma^{\rm{exp}}$. $\Sigma^{\rm{th}}$ is the sum of the theoretical uncertainties: QCD-scale, EW, and PDF\@.  The scale and EW uncertainties are fully correlated, and the PDF correlations are calculated using the Hessian prescription \cite{Butterworth:2015oua}.
$\Sigma^{\rm{stat}}$ is the statistical uncertainty which we calculate as $\Sigma^{\rm{stat}}_{ij}=\delta_{ij}\sigma_i/L$, where $L$ is the integrated luminosity. Finally, we estimate the experimental uncertainty as being the sum of uncorrelated and fully correlated components, both of which we take to be 5\% of the cross section, bin-by-bin.


{ \em Projections at 27 and 100\, TeV.---} In this Appendix we study the sensitivity of higher center of mass energy $pp$ colliders to the $Z'_Y$ model. We consider two center of mass energies: 27 and 100\,TeV. The bounds are calculated using the same procedure that we use at 13\,TeV, assuming an integrated luminosity of 1 or 10\,ab$^{-1}$. The 100\,TeV projections are compared to the constraints on the model from TLEP reach on the ${\rm Y}$ parameter, ${\rm Y}<1.5\times 10^{-4}$~\cite{Farina:2016rws}.

\begin{figure*}[!!!t]
\begin{center}
\includegraphics[width=0.9 \textwidth]{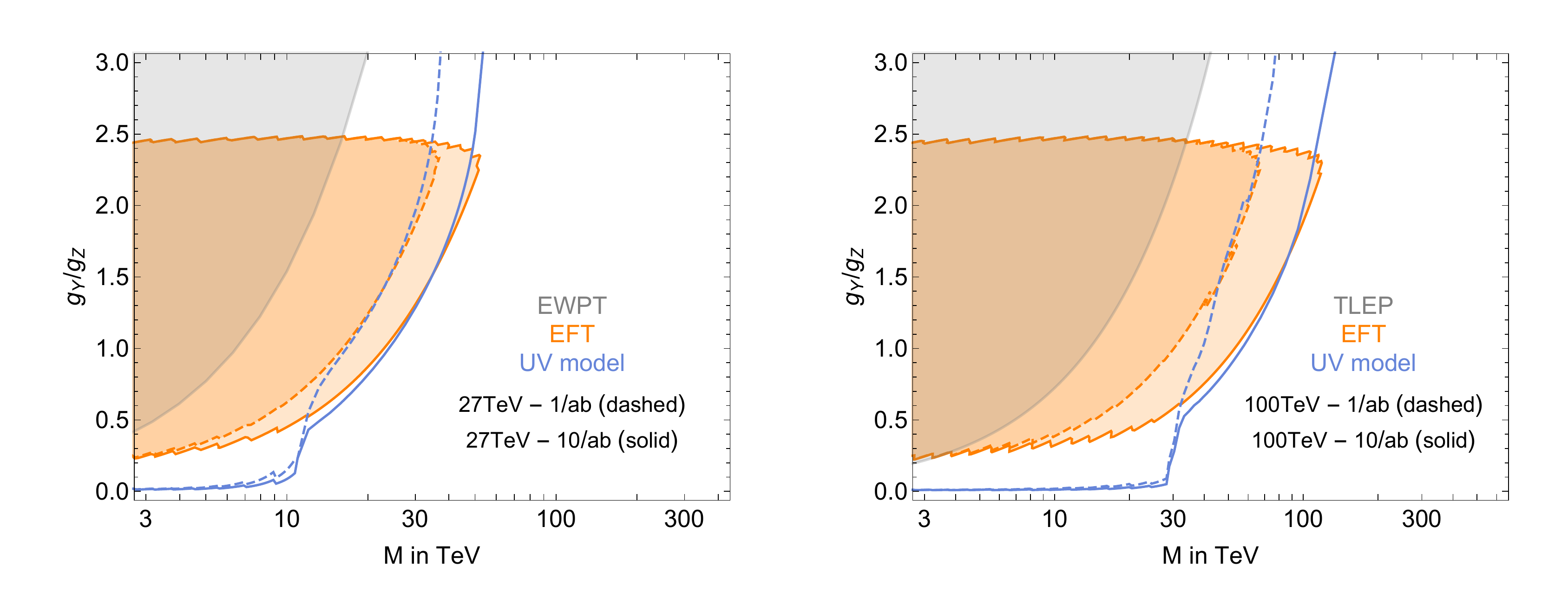}
\end{center}
\vspace{-.3cm}
\caption{\small \it
95\% C.L. bounds on the $Z'_Y$ model from $pp\to \ell^+\ell^-$, extracted using the EFT approximation (shaded orange area) and the full model (blue contours). We show projections for two different collider center of mass energies: 27\,TeV (left panel) and 100\,TeV (right panel). We compare the 27\,TeV bounds with LEP constraints on the $Z'_Y$ model, and the 100\,TeV bounds with projected TLEP limits on the ${\rm Y}$ parameter~\cite{Farina:2016rws}.
 \label{projections}}
\end{figure*}

{ \em Comparison with experimental results.---}  In this section we apply our fit to bound the scale of the following contact operators
\begin{equation}\label{4fermivalidation}
\Delta\mathscr L =\sum_{I,J\in \{L,R\}} \zeta\frac{4\pi }{\Lambda^2_{IJ}}\bar q_I\gamma_\mu q_I\, \bar \ell_J\gamma_\mu \ell_J.
\end{equation}
We sum over quarks and lepton flavors.
This is interesting as ATLAS publishes bounds on the $\Lambda_{IJ}$ in \cite{Aaboud:2017buh} using 36.1\,fb$^{-1}$ of data collected at 13\,TeV\@. Even though our fit does not correspond exactly to the experimental one (the dilepton invariant mass binning is different and experimental uncertainties with correlations are not provided by~\cite{Aaboud:2017buh}), this comparison allows us to check if our procedure leads to similar sensitivity to experimental bounds.  We find that our projections agree with the  ATLAS bounds on the operator scale within $\sim 10-30\%$, as shown in Table~\ref{valid}.

\begin{table}[htb]\label{comparison}
\renewcommand{\arraystretch}{1.}
{\small
\begin{tabular}{c|c|c}  
  &~ATLAS (TeV)~&~Our Fit (TeV)~\\ \hline \hline
{$\Lambda_{LL}$}& $22.7/30.9$ & $24.5/35.2$  \\ 
{$\Lambda_{LR}$}& $23.8/28.2$ & $26.7/31.1$  \\ 
{$\Lambda_{RL}$}& $24.0/28.0$ & $26.3/31.3$  \\ 
{$\Lambda_{RR}$} & $23.5/28.3$ & $25.1/34.6$  \\ 
\end{tabular}
}
\caption{\label{valid}\it Bound on the scale $\Lambda$ for the operators in  Eq.~\ref{4fermivalidation} obtained in \cite{Aaboud:2017buh} by the ATLAS collaboration and the limit extracted with our fitting procedure. The weaker/stronger bounds are for $\zeta$=+1/$-$1 corresponding to destructive/constructive interference with the SM\@. }
\end{table}

\bibliography{biblioZp}

\end{document}